\documentclass{iopart}
\usepackage{amssymb}
\usepackage{amsopn}

\DeclareMathOperator{\rank}{rank}
\DeclareMathOperator{\id}{Id}
\DeclareMathOperator{\idn}{\widehat{Id}}

\newtheorem{theorem}{Theorem}
\newtheorem{lemma}[theorem]{Lemma}
\newtheorem{corollary}[theorem]{Corollary}

\newenvironment{proof}{\trivlist\item[\hskip \labelsep{\textit{Proof.}}]}%
            {\hfil\hfill$\square$\endtrivlist}

\begin{document}

\title{Algebraic Rainich theory and antisymmetrisation in higher dimensions}

\author{G Bergqvist and A H\"oglund}
\address{Matematiska institutionen, Link\"opings universitet,
  SE-581 83 Link\"oping, Sweden}
\eads{\mailto{gober@mai.liu.se}, \mailto{anhog@mai.liu.se}}

\begin{abstract}
    The classical Rainich(-Misner-Wheeler) theory gives necessary
    and sufficient conditions on an energy-momentum tensor $T$ to
    be that of a Maxwell field (a 2-form) in four dimensions. 
    Via Einstein's equations these conditions can be expressed in 
    terms of the Ricci tensor, thus providing conditions on a
    spacetime geometry for it to be an Einstein-Maxwell spacetime. 
    One of the conditions is that
    $T^2$ is proportional to the metric, and it has previously
    been shown in arbitrary dimension that any tensor satisfying
    this condition is a superenergy tensor of a simple $p$-form.
    Here we examine algebraic Rainich conditions for general
    $p$-forms in higher dimensions and their relations to identities 
    by antisymmetrisation.  Using antisymmetrisation techniques we 
    find new identities for superenergy tensors of these general 
    (non-simple) forms, and we also prove in some cases the converse;
    that the identities are sufficient to determine the form. 
    As an example we obtain the complete generalisation of the 
    classical Rainich theory to five dimensions.
\end{abstract}


\maketitle

\section{Introduction}

Classical Rainich theory is concerned with studying conditions under
which a given energy-momentum tensor in four dimensions can be
constructed from a Maxwell field.  The algebraic part, which here will
be called the algebraic Rainich conditions, gives necessary and
sufficient constraints on when a given energy-momentum tensor $T_{ab}$ 
can be constructed from a 2-form, i.e., when there exists a 2-form
$F_{ab}$ such that $T_{ab}=F_{ac}F_b{}^c-\frac{1}{4}F_{cd}F^{cd}g_{ab}$.  
The conditions, first found by Rainich~\cite{rainich}, are
\begin{equation}
    T_a{}^a = 0
\end{equation}
\begin{equation}
    T_{ac}T_b{}^c \propto g_{ab}\\
\end{equation}
\begin{equation}
    T_{ab}u^av^b \geq 0\textrm{ for every pair of future-causal
    vectors $u^a$ and $v^a$}.
\end{equation}

The last condition is usually called the dominant energy condition,
and it can be replaced by some other energy condition.  Using
Einstein's equations and the fact that $T_{ab}$ is trace-free, one may
replace $T_{ab}$ by the Ricci curvature tensor $R_{ab}$ in these
equations.  Then one obtains necessary and sufficient conditions on a
four dimensional Lorentzian geometry to correspond (algebraically) to
an Einstein-Maxwell spacetime.  For $F_{ab}$ to satisfy Maxwell's
equations, one also adds a differential condition on $T_{ab}$ or
$R_{ab}$.  This view of geometrising physical situations, without
going to higher dimension than four, was especially developed by
Misner and Wheeler~\cite{misner}.  The theory is well developed for
four dimensions and 2-forms but conditions have also been developed
for massless scalar fields and some other physical situations
(see~\cite{bergqvist01} for many references).  It is of interest to
generalise the results to other cases and higher dimensions for
several reasons.  Some gravitation theories live in higher dimensions
and can therefore not make use of the classical algebraic Rainich
conditions.  It can also shed some light on the four dimensional case
by considering what happens when the number of dimensions is different
than four, thereby showing what is special to four dimensions.
Another reason is that there is some nice mathematics to be
discovered.

In Bergqvist and Senovilla~\cite{bergqvist01} superenergy tensors of
simple forms in arbitrary dimension were examined.  One of the results
found is that the square of the superenergy tensor of any simple form
is proportional to the metric.  Conversely, it was also shown that if
the square of a symmetric tensor is proportional to the metric, then
the tensor is, up to sign, the superenergy tensor of a simple
$p$-form.  Furthermore, the trace of the tensor determines the value
of $p$, i.e., what kind of form that can be used to construct the
superenergy tensor.  These results therefore give a complete
generalisation of the classical algebraic Rainich theory in the case
the condition $ T_{ac}T_b{}^c \propto g_{ab}$ is preserved.  As the
energy-momentum tensors of Maxwell fields and massless scalar fields
are special cases of superenergy tensors of $p$-forms, it seems that
superenergy tensors are the natural objects to consider when a more
general algebraic Rainich theory is sought.

In this paper we will regain these results and extend considerations
to non-simple forms.  The problem is naturally divided into two parts.
The first part is to find a necessary identity, which replaces
$ T_{ac}T_b{}^c \propto g_{ab}$, for the superenergy
tensor.  The second part is to show that this identity is the best
possible for that case, or even better, show that the condition
together with the energy condition is sufficient.

The first part will be done by antisymmetrising over the indices of
the forms involved.  In that process, and even in later cases, it
turns out that the metric structure is not relevant.  Therefore those
parts will be presented without the metric structure of the vector
space.

The metric structure is however crucial for the converse and we will
be mainly interested in real vector spaces with Lorentz signature.  In
this case we obtain complete results for 2-forms of rank 4 in
arbitrary dimension.  From this we immediately find the necessary and
sufficient conditions for an energy-momentum tensor in five dimensions
to originate in a 2-form, which is a complete generalisation of the
classical algebraic Rainich theory to five dimensions.  We also
discuss cases of higher rank of the 2-form and $p$-forms with $p\ge
3$.  However, it could also be interesting to work out the converse
for other signatures, e.g., with a positive definite metric.  The
first step would have to be to find a replacement of the dominant
energy condition.

In section~\ref{section:notation} we present some notation, basic
definitions, and a theorem for Lorentz metrics corresponding to the
spectral theorem for positive definite metrics.  In
section~\ref{section:simple} we use antisymmetrisation to study
Rainich theory for simple forms.  Then 2-forms of rank 4 are
considered in section~\ref{section:frank4}, with some of the
calculations presented in \hbox{\ref{ap:rank4sol}}.  More general forms are
treated in sections~\ref{section:highrank} and~\ref{section:pforms},
and we finish with a discussion in section~\ref{section:discussion}.

\section{Notation and prerequisites}\label{section:notation}

In this paper we will use the abstract index notation.  The letter $n$
will always refer to the number of dimensions.  With a tensor we will
mean a multilinear mapping on an $n$-dimensional vector space.  If so
preferred it can be seen as a tensor field taken pointwise on an
$n$-dimensional manifold.  When talking about $(1,1)$-tensors we will
use an index free notation.  Thus $T$ will mean $T^a{}_b$ and $T^2$
will mean $T^a{}_c T^c{}_b$ etc.  The trace $T^c{}_c$ will be written
as $[T]$.  In the context of tensors any scalar term will have an
implicit identity mapping $\delta^a_b$ on it.  Due to preference the
objects in the index free notation can instead of tensors be viewed as
linear mappings with the product being composition, or as matrices
with the product being matrix multiplication.

The metric free parts are very general.  The tensors can be taken over
any finite dimensional vector space over any field of zero
characteristic (a sufficient condition for the results in this paper).
Occasionally we need to extend the field in order to solve for
eigenvalues.  The reader who so prefers, may just think of vector 
spaces over real numbers and extend that to complex numbers 
whenever necessary.

Whenever we talk about the algebraic multiplicity of an eigenvalue we
refer to its multiplicity as a solution of the characteristic
equation, not its multiplicity as a solution of the equation at hand.
The geometrical multiplicity refers to the dimension of the eigenspace
associated with the eigenvalue.

A $p$-form $A$ is a tensor with $p$ indices that is totally
antisymmetric.  
The \emph{rank} of a $p$-form $A$ is the dimension of the subspace that is
spanned by
\begin{equation}
    A_{a_1a_2\ldots a_p} u^{a_2}v^{a_3}\dots w^{a_p}
\end{equation}
when the vectors $u^a, \ldots, w^a$ varies.  Equivalently, the rank of a
$p$-form $A$ is the lowest number of linearly independent
(dual-)vectors that must be used in order to construct the $p$-form.

The  $p$-form $A$ is  \emph{simple} if it is a product of 1-forms,
i.e.,
\begin{equation}
    A_{a_1\ldots a_p} = u_{[a_1}\dots w_{a_p]}
\end{equation}
where $u_a, \ldots, w_a$ are 1-forms, and where the brackets
$[\ ]$ denote the antisymmetric part.

The superenergy tensor $T_{ab}$ of a $p$-form $A=A_{c_1\ldots c_p}$,
$1\le p\le n$, is given by~\cite{bergqvist01,seno2000}
\begin{equation}\label{eq:origse}
    T_{ab} =
    T_{ab}\{A\}= \frac{1}{(p-1)!}\Bigl(
        A_a{}^{c_2\ldots c_p}A_{bc_2\ldots c_p}
        -\frac{1}{2p}A^{c_1\ldots c_p}A_{c_1\ldots c_p} g_{ab}\Bigr)
\end{equation}
if the metric $g_{ab}$ is assumed to have (Lorentz) signature $(- +
\ldots +)$.  As shown in~\cite{seno2000}, for $1\leq p \leq n-1$, this
expression is equivalent to the original definition first presented by
Senovilla, also in~\cite{seno2000}, which involves $A$ and its Hodge
dual.  Note that the $p$-form $A$ and its dual ($n-p$)-form have the
same superenergy tensors if $1\le p\le n-1$ ~\cite{seno2000}.
 
We would now like to have a definition that does not include the metric 
structure.  Raising the first index on $T$ makes the metric disappear 
from the definition.
\begin{equation}
    T^a{}_b = \frac{1}{(p-1)!}\Bigl(
        A^{ac_2\ldots c_p}A_{bc_2\ldots c_p}
        -\frac{1}{2p}A^{c_1\ldots c_p}A_{c_1\ldots c_p}
        \delta^a_b\Bigr).
\end{equation}
Without a metric there is no way to relate $A_{a_1\ldots a_p}$ 
to $A^{a_1\ldots a_p}$.
We use different letters for the two forms to emphasise this.  In
order to simplify things further we omit the non-zero coefficient
since that will not affect the metric free part of this paper and will
only rescale eigenvalues for the metric dependent parts.  So, in this
paper the \emph{superenergy tensor} will be defined as
\begin{equation}\label{eq:se}
    T^a{}_b =
    T^a{}_b\{A, B\} =
        A^{ac_2\ldots c_p}B_{bc_2\ldots c_p}
        -\frac{1}{2p}A^{c_1\ldots c_p}B_{c_1\ldots c_p}
        \delta^a_b.
\end{equation}

The product of two $p$-forms $A$ and $B$ will be used frequently in
this paper, so we introduce the notation
\begin{equation}
    P^a{}_b = A^{ac_2\ldots c_p} B_{bc_2\ldots c_p}.
\end{equation}
Thus the superenergy tensor can now be written
\begin{equation}
    T=P-\frac{1}{2p}[P]
\end{equation}
using index-free notation.

If the signature $(+ - \ldots -)$ is used then the omitted coefficient 
has a sign that depends on $p$ and that will of course not just add a 
rescaling but will also change the sign of the eigenvalues.

Whenever a metric is present we assume that $A$ and $B$ are equal
relative to the metric.  That implies that the superenergy tensor
is symmetric relative to this metric, i.e., $g_{ac}T^c{}_b =
g_{bc}T^c{}_a$.

The fact that $T$ is symmetric relative to a given metric says something
about the structure of $T$.  Exactly what it says depends on the
metric.  If the metric is positive definite in a real vector space
then the spectral theorem says that there is a basis of orthogonal
eigenvectors of $T$.  Here we are mainly concerned with metrics with
Lorentz signature $(- + \ldots +)$, and then the situation is
a little bit more involved.  The following theorem covers that case.
The formulation is due to Hall et al.~\cite{hall96} and is slightly
adopted to this paper and extended to arbitrary dimension.  The
Segr\'e types are not used elsewhere in this paper and can here be
considered as just a convenient classification.
\begin{theorem}\label{th:spectrallorentz}
    Let $V$ be an $n$-dimensional vector space over the real numbers,
    $n\geq 3$, with a Lorentz metric $g$ with signature $(- + \dots
    +)$.  Let $T_{ab}$ be a real symmetric second order tensor on
    $V$.  Then (with respect to $g$) $T$ must take one of the Segr\'e
    types $\{1,1111\ldots\}$, $\{2111\ldots\}$, $\{311\ldots\}$ and
    $\{z \bar z 111\ldots\}$ or a degeneracy thereof.  The
    corresponding canonical forms in some appropriate basis (either
    orthonormal $t^a$, $(x_i)^a$, $i=2\ldots n$, or null $l^a$, $m^a$,
    $(x_i)^a$, $i=3\ldots n$, in which the only non-vanishing inner
    products are respectively, $-t_at^a = (x_i)_a(x_i)^a = 1$ and
    $l_am^a = (x_i)_a(x_i)^a = 1$) are:

    $(i)$ for the type $\{1,111\ldots\}$ either of
    \begin{equation}\label{eq:symm1a}
        T_{ab} =
        \alpha_1 t_at_b
        +\sum_{i=2}^n\alpha_i (x_i)_a(x_i)_b
    \end{equation}
    or, equivalently using  $l^a$ and $m^a$,
    \begin{equation}
        T_{ab} =
        (\alpha_2-\alpha_1)l_{(a}m_{b)}
        +\case{1}{2}(\alpha_1+\alpha_2)(l_al_b+m_am_b)
        +\sum_{i=3}^n\alpha_i (x_i)_a(x_i)_b
    \end{equation}
    
    $(ii)$ for the type $\{2111\ldots\}$
    \begin{equation}
        T_{ab} =
        2\beta_1 l_{(a}m_{b)}
        \pm l_al_b
        +\sum_{i=3}^n\beta_i (x_i)_a(x_i)_b
    \end{equation}

    $(iii)$ for the type $\{311\ldots\}$
    \begin{equation}
        T_{ab} =
        2\gamma_1 l_{(a}m_{b)}
        +2 l_{(a}(x_3)_{b)}
        +\sum_{i=3}^n\gamma_i (x_i)_a(x_i)_b
    \end{equation}

    $(iv)$ finally for the type $\{z\bar z111\ldots\}$
    \begin{equation}
        T_{ab} =
        2\delta_1 l_{(a}m_{b)}
        +\delta_2 (l_al_b-m_am_b)
        +\sum_{i=3}^n\delta_i (x_i)_a(x_i)_b
    \end{equation}

    where $\alpha_1, \alpha_2, \ldots, \delta_n \in \mathbb{R}$ and
    $\delta_2\not = 0$.
\end{theorem}
Case $(i)$ means that there is a basis of eigenvectors.  In case $(iv)$
there is a basis of complex eigenvectors, which here means that there
is an invariant two dimensional subspace and real eigenvectors.  In
the other two cases there is a basis of eigenvectors and
Jordan-strings.  A Jordan-string of a linear mapping $T^a{}_b$ is a
sequence of vectors $(e_i)^a$, $i=1\dots k$ such that $T^a{}_b (e_i)^b
= \lambda (e_i)^a + (e_{i-1})^a$ for $i=2\dots k$ and $T^a{}_b (e_1)^b
= \lambda (e_1)^a$.  If $k$ is maximal then $k$ is said to be the
length of the Jordan-string.  If the length is one then vi call the
Jordan-string trivial.

The fact that the forms $A$ and $B$ are equal may under some circumstances
imply further conditions.  In the case of Lorentz metric in a real
vector space the following general result holds for any $T$ defined
by~\eref{eq:origse}
\begin{equation}
    T_{ab}u^av^b \geq 0\textrm{ for every pair of future-causal
    vectors $u^a$ and $v^a$}.
\end{equation}
As stated above, this is the dominant energy condition (DEC).  It is a
special case of the so-called dominant property which has been proven
for general superenergy tensors with an arbitrary number of
indices~\cite{bergqvist99,seno2000}.

It is straightforward to check which cases in 
Theorem~\ref{th:spectrallorentz} that are consistent with the DEC.
In case $(i)$ we shall in what follows only need $T$ on the
form~\eref{eq:symm1a}.

\begin{theorem}\label{th:spectrallorentzdec}
    Assumptions as in Theorem~\ref{th:spectrallorentz}.  If, in
    addition, the DEC is satisfied then the following cases are the
    only possible ones.
    
    $(i)$ for the type $\{1, 111\ldots\}$ 

    \begin{equation}
        T_{ab} =
        \alpha_1 t_at_b
        +\sum_{i=2}^n\alpha_i (x_i)_a(x_i)_b
    \end{equation}

    $(ii)$ for the type $\{2111\ldots\}$
    \begin{equation}
        T_{ab} =
        2\beta_1 l_{(a}m_{b)}
        + l_al_b
        +\sum_{i=3}^n\beta_i (x_i)_a(x_i)_b
    \end{equation}
    where $-\alpha_1 \leq \alpha_i \leq \alpha_1$ and 
    $\beta_1 \leq \beta_i \leq -\beta_1$.
\end{theorem}

\section{Simple forms}\label{section:simple}

\subsection{Metric free}

\begin{lemma}\label{le:simple}
    Let $P^a{}_b = A^{ac_2\ldots c_p}B_{bc_2\ldots c_p}$ where $A$ and
    $B$ are $p$-forms and $A$ is simple.  Then 
    \begin{equation}
        0 = P\Bigl(P-\frac{1}{p}[P]\Bigr).
    \end{equation}
\end{lemma}
\begin{proof}
    Using the property that $A$ is simple gives the first equality
    \begin{eqnarray}
        \fl 0 = (p+1) A^{[c_1\ldots c_p}A^{a]d_2\ldots d_p}
        B_{c_1\ldots c_p}B_{bd_2\ldots d_p}
        \nonumber\\
        \lo =
        A^{c_1\ldots c_p}B_{c_1\ldots c_p}
        A^{ad_2\ldots d_p}B_{bd_2\ldots d_p}
        \nonumber \\
        -p A^{ac_2\ldots c_p}B_{c_1c_2\ldots c_p}
        A^{c_1d_2\ldots d_p}B_{bd_2\ldots d_p}.
    \end{eqnarray}
    Translating this into index free notation gives the lemma.
\end{proof}
\begin{theorem}\label{th:simple}
    If $A$ and $B$ are $p$-forms and $A$ is simple  then
    $T=T^a{}_b\{A,B\}$ satisfies
    \begin{equation}
        T^2=\frac{1}{(2p)^2}[P]^2
    \end{equation}
    which, if $n\not=2p$, can be written as
    \begin{equation}\label{eq:t2}
        T^2=\frac{1}{(n-2p)^2}[T]^2.
    \end{equation}
\end{theorem}
This is a more general result than the one presented in~\cite{bergqvist01}
as only $A$ is assumed to be simple.

The following Corollary follows by repeatedly multiplying
equation~\eref{eq:t2} by $T$ and taking traces.
\begin{corollary}
    If $A$ and $B$ are $p$-forms, $n\not=2p$ and $A$ is simple  then
    $T=T^a{}_b\{A,B\}$ satisfies
     \begin{equation}
         [T^k] =
         \cases{
           \frac{n}{(n-2p)^k}[T]^k      & if $k$ is even\\
           \frac{1}{(n-2p)^{k-1}}[T]^k  & if $k$ is odd.
         }
     \end{equation}
\end{corollary}

\subsection{Lorentz metric}

We now assume that $T$ is symmetric, and satisfies $T^2 = f$ and the DEC.

That $T$ satisfies $T^2=f$ gives that $T$ have the two eigenvalues $\pm\sqrt{f}$.
Theorem~\ref{th:spectrallorentzdec} shows that these must be real,
which means that $f\geq 0$.

If $f=0$ then either we have case $(i)$ in Theorem~\ref{th:spectrallorentzdec}
with all coefficients zero implying $T=0$, or we have case $(ii)$ with
all coefficients zero.  The 1-form $l_a$ gives this $T$.  The $p$-form
obtained from $l_a$ and $p-1$ vectors orthogonal to $l_a$ also gives
this $T$.

If $f\not=0$ then we have case $(i)$ with $\alpha_1 = +\sqrt{f}$ and $\alpha_i=\pm\sqrt{f}$ for
$i=2\ldots n$.  This means that we have two non-degenerate
eigenspaces: a space-like eigenspace with eigenvalue $+\sqrt{f}$ and an
eigenspace containing a time-like direction with eigenvalue
$-\sqrt{f}$.  
Constructing a form from a basis of any of these two eigenspaces and 
scaling it appropriately gives this particular $T$.

If $k$ is the dimension of the eigenspace containing the time-like
direction, then $[T]=(n-2k)\sqrt{f}$; the proportionality constant 
$f$ and the trace of $T$ give the possible values on $p$, i.e., 
$p=k$ or $p=n-k$.

We have thus proved of the following  theorem.

\begin{theorem}
    A real symmetric tensor $T$ of a real vector space with signature $(- +
    \dots +)$ is the superenergy tensor of a simple form  if and
    only if

    a) $T$ satisfies DEC

    b) $T^2 = f$
    
    Further, if $f=0$ then the form is a null-form and if $f\not=0$ then
    the form can be chosen as a space-like
    $\frac{1}{2}(n-[T]/\sqrt{f})$-form or as a 
    $\frac{1}{2}(n+[T]/\sqrt{f})$-form containing a
    time-like direction.
\end{theorem}
This theorem was first proved in~\cite{bergqvist01}, but the proof
presented here is both simpler and shorter.  In the next section we
will use our new methods to generalise the theorem to cases with
non-simple forms.

\section{2-forms of rank four}\label{section:frank4}

\subsection{Metric free}

\begin{theorem}\label{th:rank4}
    Let $A$ and $B$ be 2-forms, $\rank A \leq 4$, $n\not=4$ and
    $T=T^a{}_b\{A,B\}$.  Then
    \begin{equation}\label{eq:rank4}
        \Bigl( T^2 -\frac{1}{4}[T^2]+\frac{1}{4(n-4)}[T]^2 \Bigr)
        \Bigl( T-\frac{1}{n-4}[T] \Bigr)
        = 0.
    \end{equation}
\end{theorem}
\begin{proof}
    Since $A$ has rank at most four the following holds,
      \begin{equation}
        0=A^{a[c}A^{de}A^{fg]}B_{bc}B_{de}B_{fg}.
    \end{equation}
    Expanding the antisymmetry and collecting the forms together
    appropriately gives
    \begin{equation}
        0 =
        P^3
        -\case{1}{2}[P]P^2
        -\case{1}{4}[P^2]P
        +\case{1}{8}[P]^2P.
    \end{equation}
    It is possible to rewrite this with $T$ instead of $P$ provided that the
    number of dimensions is not four,
    \begin{eqnarray}
        \fl 0 =
        T^3
        -\frac{1}{n-4}[T]T^2
        -\frac{1}{4}[T^2]T
        +\frac{1}{4(n-4)}[T]^2T
        \nonumber\\
        +\frac{1}{4(n-4)}[T^2][T]
        -\frac{1}{4(n-4)^2}[T]^3.
    \end{eqnarray}
    Factoring this gives the theorem.
\end{proof}

Solving equation~\eref{eq:rank4} is a little bit more involved than
the previous case; see \hbox{\ref{ap:rank4sol}} for details.
Table~\ref{tab:solutions} summarises the solutions in terms of
eigenvalues and Jordan strings for the product $P$ and the superenergy
tensor $T$ (which is just a shift in the eigenvalues but otherwise the
same structure).  While solving the equation it might be necessary to
extend the field of numbers in order for the square roots to exist.

\begin{table}[htb]
    \caption{Solutions of equation~\eref{eq:rank4}}
    \label{tab:solutions}
    \begin{tabular}{cccc}
            \br
            &&Alg.& Max.~length\\
            \centre{1}{Eigenvalues of $T$} &
            \centre{1}{Eigenvalues of $P$} &
            mult. & Jordan-str.\\
            \mr
            \multicolumn{4}{@{}l}{Case a:
              $\displaystyle{
                \frac{1}{n-4}[T]^2
                \not =
                [T^2]
                \not =
                \frac{n}{(n-4)^2} [T]^2
              }$}
            \\
            \mr
            $\displaystyle{+\frac{1}{2}\sqrt{[T^2]-\frac{1}{n-4}[T]^2}}$
            &
            $\displaystyle{-\frac{1}{n-4}[T]
              + \frac{1}{2}\sqrt{[T^2]-\frac{1}{n-4}[T]^2}}$
            &
            2 & 1
            \\
            $\displaystyle{-\frac{1}{2}\sqrt{[T^2]-\frac{1}{n-4}[T]^2}}$
            &
            $\displaystyle{-\frac{1}{n-4}[T]
              - \frac{1}{2}\sqrt{[T^2]-\frac{1}{n-4}[T]^2}}$
            &
            2 & 1
            \\
            $\displaystyle{\frac{1}{n-4}[T]}$&
            0 &
            $n-4$ & 1
            \\
            \mr
            \multicolumn{4}{@{}l}{Case b:
              $\displaystyle{
                \frac{1}{n-4}[T]^2
                =
                [T^2]
                \not =
                \frac{n}{(n-4)^2} [T]^2
              }$}
            \\
            \mr
            0 &
            $\displaystyle{-\frac{1}{n-4}[T]}$&
            4 & 2
            \\
            $\displaystyle{\frac{1}{n-4}[T]}$&
            0 &
            $n-4$ & 1
            \\
            \mr
            \multicolumn{4}{@{}l}{Case c:
              $\displaystyle{
                \frac{n}{(n-4)^2} [T]^2
                =
                [T^2]
                \not =
                0
              }$}
            \\
            \mr
            $\displaystyle{-\frac{1}{n-4}[T]}$&
            $\displaystyle{-\frac{2}{n-4}[T]}$&
            2 & 1
            \\
            $\displaystyle{\frac{1}{n-4}[T]}$&
            0 &
            $n-2$ & 2
            \\
            \mr
            \multicolumn{4}{@{}l}{Case d:
              $\displaystyle{
                [T] = 0 = [T^2]
              }$}
            \\
            \mr
            0 & 0 & $n$ & 3
            \\
            \br
    \end{tabular}
\end{table}

\subsection{Lorentz metric}

Assume now that we have a vector space over the real numbers with a
metric with Lorentz signature, and that $T$ is symmetric with respect
to this metric and satisfies the DEC and equation~\eref{eq:rank4}.
Then the assumptions in Theorem~\ref{th:spectrallorentzdec} are
satisfied and we can compare the solutions given above with the two
cases in the theorem.  We want to show that if these conditions are
satisfied then there are forms that gives this particular superenergy
tensor.

Solution (a) corresponds to case (i) in the theorem.  All eigenvalues
are real so $[T^2]-[T]^2/(n-4)>0$.  Then there are two cases:
$[T^2] > n[T]^2/(n-4)^2$ and $[T^2] < n[T]^2/(n-4)^2$.  In
the first case
$\frac{1}{2}\sqrt{[T^2]-[T]^2/(n-4)}>\left|[T]/(n-4)\right|$ which
means that the timelike eigenvalue is
$-\frac{1}{2}\sqrt{[T^2]-[T]^2/(n-4)}$.  Numbering the other vectors in
the theorem such that $(x_2)^a$ has the same eigenvalue as the
timelike vector and $(x_3)^a$ and $(x_4)^a$ have the other eigenvalue
with multiplicity two, then we can find that one form that gives this
superenergy tensor is
\begin{eqnarray}
    \fl 2\sqrt{\frac{1}{n-4}[T]+\frac{1}{2}\sqrt{[T^2]-\frac{1}{n-4}[T]^2}}\;
        t_{[a} (x_2)_{b]}
    \nonumber \\
    +2\sqrt{-\frac{1}{n-4}[T]+\frac{1}{2}\sqrt{[T^2]-\frac{1}{n-4}[T]^2}}\;
        (x_3)_{[a} (x_4)_{b]}.
\end{eqnarray}

In the other case, $[T^2] < n[T]^2/(n-4)^2$, the theorem gives
$\frac{1}{2}\sqrt{[T^2]-[T]^2/(n-4)} < \left|[T]/(n-4)\right|$ which
means that the timelike eigenvalue must be $[T]/(n-4)$ and that
that $[T]$ must be negative.  Numbering the vectors in the theorem
appropriately showes that $T$ is obtained from
\begin{eqnarray}
    \fl 2\sqrt{-\frac{1}{n-4}[T]+\frac{1}{2}\sqrt{[T^2]-\frac{1}{n-4}[T]^2}}\;
        (x_2)_{[a} (x_3)_{b]}
    \nonumber \\
    +2\sqrt{-\frac{1}{n-4}[T]-\frac{1}{2}\sqrt{[T^2]-\frac{1}{n-4}[T]^2}}\;
        (x_4)_{[a} (x_5)_{b]}.
\end{eqnarray}
Solution (b) in the case where there are only trivial Jordan-strings
gives case (i) in the theorem.  The timelike eigenvalue must be
$[T]/(n-4)$ and $[T]$ must be negative.  One form that gives
this superenergy tensor is
\begin{equation}
    2\sqrt{-\frac{1}{n-4}[T]}\;
        \Bigl((x_2)_{[a} (x_3)_{b]} + (x_4)_{[a} (x_5)_{b]}\Bigr).
\end{equation}

Solution (b) in the case where there are non-trivial Jordan-strings
corresponds to case (ii) in the theorem.  However, the eigenvalue
$\beta_1$ that permits Jordan-strings in solution (b) is zero 
and that is not compatible with case (ii) in the theorem, i.e.,
the DEC is not satisfied.

Solution (c) in the case where there are only trivial Jordan-strings
gives case (i) in the theorem.  The timelike eigenvalue is
$[T]/(n-4)$ if $[T]$ is negative and $-[T]/(n-4)$ if
$[T]$ is positive.  In the first case the form
\begin{equation}
    2\sqrt{-\frac{2}{n-4}[T]}\;
        t_{[a} (x_2)_{b]}
\end{equation}
gives this superenergy tensor and in the second case
\begin{equation}
    2\sqrt{\frac{2}{n-4}[T]}\;
        t_{[a} (x_2)_{b]}
\end{equation}
gives it.

Solution (c) in the case where there is a non-trivial Jordan-string
gives case (ii) in the theorem.  Then $[T]$ must be negative and
$[T]/(n-4)$ is the null eigenvalue.  The form
\begin{equation}
    2l_{[a} (x_3)_{b]}
    +2\sqrt{-\frac{2}{n-4}[T]}\;
        (x_4)_{[a} (x_5)_{b]}
\end{equation}
gives this $T$.

Solution (d) in the case where there are only trivial Jordan-strings
gives that $T=0$.

Solution (d) in the case where there are non-trivial Jordan-strings
gives case (ii) in the theorem (which excludes Jordan-strings of
length three) where all $\beta_i$ are zero.  The form
\begin{equation}
        2l_{[a} (x_2)_{b]}
\end{equation}
gives this superenergy tensor.

Thus we have proved the following theorem.
\begin{theorem}\label{th:rank4Rainich}
    A real symmetric tensor $T=T^a{}_b$ in an $n$-dimensional real
    vector space, $n > 4$, with Lorentz signature is the superenergy
    tensor of a 2-form of rank at most four if and only if
    
    a) $T$ satisfies DEC
    
    b) $\displaystyle{\Bigl( T^2 -\frac{1}{4}[T^2]+\frac{1}{4(n-4)}[T]^2 \Bigr)
    \Bigl( T-\frac{1}{n-4}[T] \Bigr) = 0}$.
    
    Further, if $(n-4)^2[T^2]<n [T]^2$ then the 2-form is the sum of
    two simple spacelike 2-forms, and if $(n-4)^2[T^2]>n [T]^2$ then
    the 2-form is the sum of one simple spacelike 2-form and one simple
    2-form containing a timelike direction.
\end{theorem}

Since the dual of a $p$-form gives the same superenergy
tensor as the $p$-form itself we have the following corollary.
\begin{corollary}
    A real symmetric tensor $T=T^a{}_b$ in an $n$-dimensional real
    vector space, $n > 4$, with Lorentz signature is the superenergy
    tensor of a ($n-2$)-form which is a sum of at most two simple
    ($n-2$)-forms if and only if
    
    a) $T$ satisfies DEC
    
    b) $\displaystyle{\Bigl( T^2 -\frac{1}{4}[T^2]+\frac{1}{4(n-4)}[T]^2 \Bigr)
    \Bigl( T-\frac{1}{n-4}[T] \Bigr) = 0}$.
    
    Further, if $(n-4)^2[T^2]<n [T]^2$ then the ($n-2$)-form is the sum of
    two simple ($n-2$)-forms containing a timelike direction, and if 
    $(n-4)^2[T^2]>n [T]^2$ then the ($n-2$)-form is the sum of one 
    simple spacelike ($n-2$)-form and one simple ($n-2$)-form
    containing a timelike direction.
\end{corollary}

\subsection{Five dimensions, Lorentz metric}

All cases in five dimensional spaces with a Lorentz metric are now
covered; 1-forms and 4-forms are always simple, 2-forms cannot have
higher rank than four and 3-forms can be referred back to 2-forms by
taking duals of them.  Hence, an immediate consequence of
Theorem~\ref{th:rank4Rainich} is
\begin{theorem}
    A real symmetric tensor $T=T^a{}_b$ in an $5$-dimensional real
    vector space with Lorentz signature is the superenergy tensor of a
    2-form if and only if
    
    a) $T$ satisfies DEC
    
    b) $\displaystyle{\Bigl( T^2 -\frac{1}{4}[T^2]+\frac{1}{4}[T]^2 \Bigr)
      \Bigl( T-[T] \Bigr) = 0}$.
\end{theorem}
This theorem is the generalisation of the classical algebraic Rainich
theory to five dimensions.  It gives necessary and sufficient
conditions on $T$ to correspond algebraically to the energy-momentum
tensor of an electromagnetic field.  Note that the difference with
four dimensions is the factor $T-[T]$, and that the trace of $T$,
which is non-zero in five dimensions, is built into the equation.

\subsection{Four dimensions, metric free}

Four dimensions were excluded in Theorem~\ref{th:rank4}.  That case
was already discussed in detail in Edgar and
H\"oglund~\cite{edgar2002}.  It turns out that the interesting
identity is a dimensionally dependent identity.  Antisymmetrising over
five indices, one more than the number of dimensions, gives the
identity
\begin{equation}
    0 = \delta^{[a}_b A^{cd} A^{ef]} B_{cd} B_{ef}.
\end{equation}
Expanding the antisymmetry and substituting for $T$ gives the
following theorem.
\begin{theorem}
    If $A$ and $B$ are 2-forms in four dimensions then $T=T^a{}_b\{A,B\}$
    satisfies
    \begin{equation}
        T^2=\case{1}{4}[T^2].
    \end{equation}
\end{theorem}
Combining this with Theorem~\ref{th:simple} shows that the square of
all superenergy tensors of forms are proportional to the
identity mapping in four dimensions.

Note that also Lovelock~\cite{lovelock} used antisymmetrising
and two different forms $A$ and $B$ to study this case.

\section{2-forms of higher rank}\label{section:highrank}

It is always possible to get an identity for 2-forms of at most a
given rank $k$ by antisymmetrising over $k+1$ indices in the following
way.
\begin{equation}\label{eq:2formrankhigh}
    0=A^{a[c_1}A^{c_2c_3}\dots A^{c_kc_{k+1}]}
      B_{bc_1}B_{c_2c_3}\dots B_{c_kc_{k+1}}.
\end{equation}
This will give an identity of order $k/2+1$ (note that the rank of a 2-form
is always even).

In the special case where $n=k$ we can get a dimensionally dependent
identity of order $k/2$ by antisymmetrising over $n+1$ indices in the
following way.
\begin{equation}\label{eq:2formrankhighn}
    0 = \delta^{[a}_b
    A^{c_1c_2} \dots A^{c_{k-1}c_k]}
    B_{c_1c_2} \dots B_{c_{k-1}c_k}.
\end{equation}

It is not possible to get a lower order identity for any of these
cases.  To see this, take $k$ linearly independent vectors $(v_i)^a$
and $k$ dual vectors $(\omega^i)_a$ such that
\begin{equation}
    (\omega^i)_a (v_j)^a =
    \cases{
      1 & if $i=j$\\
      0 & if $i\not = j$
    }
\end{equation}

and define 2-forms  $A$ and $B$ by
\begin{eqnarray}
    A^{ab}=\alpha_1(v_1)^{[a}(v_2)^{b]}+\dots 
    +\alpha_{k/2}(v_{k-1})^{[a}(v_k)^{b]}
    \\
    B_{ab}=\alpha_1(\omega^1)_{[a}(\omega^2)_{b]}+\dots
    +\alpha_{k/2}(\omega^{k-1})_{[a}(\omega^k)_{b]}
\end{eqnarray}
where all $\alpha_j^2$, which will be the eigenvalues of the
product $P$, are chosen distinct.  In the first case $P$ will also
have the eigenvalue zero.  Thus, in the two cases $P$ has $k/2+1$ 
and $k/2$ different eigenvalues respectively.  So will $T$ and
therefore cannot satisfy any polynomial equation of lower order.

When a metric is present an orthonormalised basis can of course 
be used instead of the basis presented above.

It is possible to obtain the identities~\eref{eq:2formrankhigh}
and~\eref{eq:2formrankhighn} recursively.  The following theorem tells
us how to do that.

\begin{theorem}\label{th:Idk}
    Let
    \begin{equation}
        \idn_k =
        \delta^{[a}_b
        A^{c_1c_2} \dots A^{c_{k-1}c_k]}
        B_{c_1c_2} \dots B_{c_{k-1}c_k}
    \end{equation}
    and
    \begin{equation}
        \id_k =
        A^{a[c_1}A^{c_2c_3}\dots A^{c_kc_{k+1}]}
        B_{bc_1}B_{c_2c_3}\dots B_{c_kc_{k+1}}.
    \end{equation}
    Then the following identities hold.
    \begin{equation}\label{eq:idn}
        \idn_k
        =
        -\frac{k}{k+1}
        \Bigl(
            \id_{k-2} -\frac{1}{k} [\id_{k-2}]
        \Bigr)
    \end{equation}
    \begin{equation}\label{eq:idp}
        \id_k
        =
        -\frac{k}{k+1} P
        \Bigl(
            \id_{k-2} -\frac{1}{k} [\id_{k-2}]
        \Bigr)
    \end{equation}
    \begin{equation}\label{eq:idt}
       \fl \id_k
        =
        -\frac{k}{k+1}
        \Bigl(
            T\id_{k-2}
            -\frac{1}{n-4} [T] \id_{k-2}
            -\frac{1}{k} [\id_{k-2}] T
            +\frac{1}{k(n-4)} [T][\id_{k-2}]
        \Bigr)
    \end{equation}
\end{theorem}
\begin{proof}
    Consider
    \begin{equation}
        \idn_k
        =
        \delta^{[a}_b
        A^{c_1c_2} \dots A^{c_{k-1}c_k]}
        B_{c_1c_2} \dots B_{c_{k-1}c_k}
    \end{equation}
    and split the right hand side into the two cases where in the
    first case the index $a$ is on the delta and in the second case
    $a$ is not on the delta.
    \begin{eqnarray}
        \fl (k+1) \idn_k
        =
        \delta_b^a
        A^{[c_1c_2} \dots A^{c_{k-1}c_k]}
        B_{c_1c_2} \dots B_{c_{k-1}c_k}
        \nonumber \\
        -k
        A^{[ac_2} \dots A^{c_{k-1}c_k]}
        B_{bc_2} \dots B_{c_{k-1}c_k}.
    \end{eqnarray}
    By using
    \begin{equation}\label{eq:antisymmless}
        A^{[c_1c_2} \dots A^{c_{k-1}c_k]}
        =
        A^{c_1[c_2} \dots A^{c_{k-1}c_k]}
    \end{equation}
    we get
    \begin{eqnarray}
        \fl (k+1) \idn_k
        =
        \delta_b^a
        A^{c_1[c_2} \dots A^{c_{k-1}c_k]}
        B_{c_1c_2} \dots B_{c_{k-1}c_k}
        \nonumber \\
        - k
        A^{a[c_2} \dots A^{c_{k-1}c_k]}
        B_{bc_2} \dots B_{c_{k-1}c_k}
    \end{eqnarray}
    which is the same as the identity~\eref{eq:idn}.

    The next identity is obtain in a similar way, by
    starting with
    \begin{equation}
        \id_k
        =
        A^{a[c_1}A^{c_2c_3}\dots A^{c_kc_{k+1}]}
        B_{bc_1}B_{c_2c_3}\dots B_{c_kc_{k+1}}
    \end{equation}
    and spliting the right hand side into the two cases where the index
    $c_1$ is on the first form and the case where it is not on the
    first form.
    \begin{eqnarray}
        \fl (k+1)\id_k
        =
        A^{ac_1} B_{bc_1}
        A^{[c_2c_3} \dots A^{c_kc_{k+1}]}
        B_{c_2c_3} \dots B_{c_kc_{k+1}}
        \nonumber \\
        -k A^{ac_3}B_{c_2c_3}
        A^{[c_2c_1} A^{c_4c_5} \dots A^{c_kc_{k+1}]}
        B_{bc_1} \dots B_{c_kc_{k+1}}.
    \end{eqnarray}
    Using identity~\eref{eq:antisymmless} gives identity~\eref{eq:idp}.
    
    By substituting $P=T-[T]/(n-2p)$ in
    identity~\eref{eq:idp} we obtain identity~\eref{eq:idt}.
\end{proof}
The coefficient $-\frac{k}{k+1}$ can safely be ignored if the
theorem is used to examine the identities $\id_k = 0$ and $\idn_k =
0$.  The expressions $\id_k$ and $\idn_k$ can of course be expressed
with $P$ or $T$ as preferred.

The theorem can be used for practical calculations in obtaining
identities recursively for 2-forms of higher rank, where the identity for
simple forms in Lemma~\ref{le:simple} or Theorem~\ref{th:simple} can
be used as a starting point or even go one step further back to $\id_0
= P = T-[T]/(n-4)$.

It also helps us to understand the structure of the identities for
2-forms.  The interesting case for the identity~\eref{eq:idn} is when
$k=n$.  Then it is clear from  Theorem~\ref{th:Idk} that the left hand 
side of the identity $\idn_n = 0$, given by~\eref{eq:idn}, is trivially 
trace free, i.e., it is not possible to express $[T^{n/2}]$ in other 
traces by taking the trace of this identity.  The left hand side of 
this identity will contain a term $[T^{n/2}]/n$ which
prevents us from factorising the identity.

The identity~\eref{eq:idn} also helps us understand why rank four
2-forms did not turn out to be a special case in four dimension.  We
can see that the recursion only changes the coefficient of the
constant term when going to rank $n$.  For four dimensions that did not
change the structure of the identity; the factorisation merely relied on
the coefficient to be positive (in the real case).  In higher
dimensions, we will have higher order polynomials and a change in the
constant term will have a larger impact on the factorisation.

\section{$p$-forms, $3 \leq p \leq n-3$}\label{section:pforms}

The method of antisymmetrising over the indices of the forms gets more
complicated for $p$-forms when $p \geq 3$.  If we antisymmetrise over
at least two indices on one form and over at least three forms then
it happens that the indices that should contract with the
indices of one form are spread out over three or more forms.  Then it
is not possible to combine the forms in such a way that they can be
written in terms of the product $P$.

Other possibilities fail in the same way, i.e., the forms cannot be
combined so that they can be written in terms of the product $P$.  The
best way we can do is to antisymmetrise over one index on each form.
This can be written in terms of $P$ directly as
\begin{equation}\label{eq:p3}
    0 = P^{[a}{}_b P^{c_1}{}_{c_1} \dots P^{c_k]}{}_{c_k}
\end{equation}
where the form $A$ in the product $P$ is of rank at most $k$.  For the special
case when $n=k$ we can get one order lower by
\begin{equation}\label{eq:p3lower}
    0 = \delta^{[a}_b P^{c_1}{}_{c_1} \dots P^{c_k]}{}_{c_k}
\end{equation}which just gives us the Cayley-Hamilton equation.

This complication is not just a break down of the method of getting
the identities by antisymmetrising but, as the following example
shows, a complication in the structure of the product of the forms.

Assume we have a vector space over the real numbers and a metric with
Lorentz signature or a positive definite metric.  Let $(e_i)^a$,
$i=1\ldots n$ be an orthonormalised basis with $(e_1)^a$ as the
timelike base vector in the Lorentz case.  Take the vectors
\begin{eqnarray}
    x^a = (e_1)^a + (e_2)^a \quad&
    y^a = (e_3)^a + 2(e_4)^a \quad&
    z^a = (e_5)^a + 3(e_6)^a
    \nonumber \\
    u^a = (e_1)^a - (e_2)^a&
    v^a = (e_3)^a - 2(e_4)^a&
    w^a = (e_5)^a - 3(e_6)^a
\end{eqnarray}
and build the form
 \begin{equation}
     A^{abc}=x^{[a}y^bz^{c]}+u^{[a}v^bw^{c]}.
\end{equation}
Use the metric to lower the indices on $A$ to get $B$.  The base vectors
$(e_i)^a$, $i=1\dots6$, are all eigenvectors of the product $P^a{}_b =
A^{acd}B_{bcd}$ with distinct non-zero eigenvalues.  If the dimension
is more than six then $(e_i)^a$, $i=7\ldots n$, are eigenvectors with
eigenvalue zero.  Thus, this $P$, and therefore the corresponding $T$,
can not satisfy a polynomial identity of lower order than six
respectively seven.

This example does of course not rule out the possibility of special
cases where it might be possible to obtain something better than the
identities~\eref{eq:p3} and~\eref{eq:p3lower}.

\section{Discussion}\label{section:discussion}

The main point in this paper is to highlight the relation between
identities from antisymmetrisation and Rainich theory.  We have shown
how to obtain such identities in several cases and in some cases also
shown that they are sharp in the sense that they, together with the
DEC, are sufficient conditions for those cases.  In some other cases
we have argued that there is no better algebraic identity without
showing that everything which satisfies the particular identity is
also a superenergy tensor of that type.

As we have seen in this paper the problem naturally divides into two
parts: $(i)$ obtaining identities, and $(ii)$ showing that the identity so
obtained is the best possible or, even stronger, show that it
characterises that class of superenergy tensors.

It is interesting that the first part (if the energy condition is not
included) is independent of the metric.  The
generalisations come ``for free'' once this fact is respected.
Nothing would have been simpler by restricting to real vector spaces with
some special metric and assuming that $A$ and $B$ were equal with
respect to that metric.  Actually, the absence of a metric prevents us
from unnecessary raising or lowering indices and thereby helps us to
avoid complicating the derivations unnecessarily.

The antisymmetrisations that are the foundations for the identities
have different roles in the identities obtained in this paper.  In most
cases the antisymmetrisation is over more indices than the rank of the
tensors involved.  On some occasions the number of indices that is
antisymmetrised over is larger than the number of dimensions and the
identities can thus be viewed as dimensionally dependent identities.  
When this happens it is not important that the indices 
antisymmetrised over belong to a particular tensor but can be on any 
tensor.  Normally this is used to replace one pair of forms by a delta and
thus reducing the order of the final identity.  However, when the rank
of the tensors involved is strictly less than the number of dimensions
there is nothing to gain from antisymmetrising over more indices just
to get a dimensionally dependent identity.

It is surprising that it is only the structure of one of the forms $A$
and $B$, the simpler one if given a choice, that matters.  The forms
always have the same structure when constructing superenergy
tensors so it is hard to see that this will have any practical impact.
However, from a mathematical point of view it is interesting.

A natural question to ask is if it is possible to find forms $A$ and
$B$ for any tensor $T$ that satisfies any of the identities obtained
in this paper.  In other words, does the energy condition together
with the symmetry assumption correspond to the requirement that $A$
and $B$ should be equal with respect to a given metric?  It is obvious
that we get more tensors $T$ when $A$ and $B$ are aloud to be
independent.  One way to see this is to take any $T$ that satisfies
the energy condition and then change sign on one of the forms.  The
new superenergy tensor so obtained does not satisfy the energy
condition except in special cases.

However, there are tensors $T$ which satisfy an identity without
being obtainable as a superenergy tensor.  One such tensor in five
dimensions is the one with a Jordan-string of length three and all
eigenvalues zero.  The fact that the trace is zero implies that the
tensor $T$ and the product of the forms are identical.  This tensor 
satisfies the
identity in Theorem~\ref{th:rank4} but it does not satisfy the identity
in Theorem~\ref{th:simple} so neither $A$ nor $B$ can be simple.  This
means that the forms must be 2-forms or 3-forms.  Those two cases are
equivalent in five dimensions so we may assume 2-forms, which is also
what we would expect since Theorem~\ref{th:rank4} refers to 2-forms.
Since the 2-forms cannot be simple they must both be of rank~4 which
means that the dimension of the kernel is one.  The kernel of the
composition of these two forms, and thus the tensor $T$, has at most
dimension two.  This is in contradiction with the assumed $T$ whose
kernel has dimension three.

Another question which comes naturally is whether we could have
avoided the DEC in Theorem~\ref{th:rank4Rainich} by allowing that
either of $\pm T$ is a superenergy tensor as was done
in~\cite{bergqvist01}.  This is not the case here as is easy to see by
considering, e.g., the solution of equation~\eref{eq:rank4} where all
eigenvalues are zero and with one Jordan-string of length three.  This
solution is compatible with a symmetric $T$ according to
Theorem~\ref{th:spectrallorentz}.  However, neither of $\pm T$ is
compatible with the DEC according to
Theorem~\ref{th:spectrallorentzdec}.

We have seen that the complexity of the algebraic Rainich theory
increases with increasing dimension $n$, and increasing rank of the
$p$-forms.  Furthermore, we have not discussed the non-uniqueness
problem; how many forms can have the same superenergy tensor, a
problem which has been solved in four dimensions (see,
e.g.,~\cite{bergqvist01}).  There are indications that there is a
higher degree of uniqueness if $n\ne 2p$ than if $n=2p$.  One may also
develop a Rainich theory for superenergy tensors with more indices,
such as the Bel and Bel-Robinson tensors~\cite{seno2000}.

\ack

G.B.~is grateful to the (former) Ministerio de Educaci\'on, Cultura y
Deporte in Spain for support for a sabatical year 2001 at the
University of the Basque Country in Bilbao (SAB1999-0135).

\appendix

\section{Solving equation~\eref{eq:rank4}}\label{ap:rank4sol}

The equation we want to solve is
$$
    \Bigl( T^2 -\frac{1}{4}[T^2]+\frac{1}{4(n-4)}[T]^2 \Bigr)
    \Bigl( T-\frac{1}{n-4}[T] \Bigr)
    = 0.
    \eqno \hbox{\eref{eq:rank4}}
$$

It is clear that we have at most three eigenvalues
\begin{equation}
    \pm\frac{1}{2}\sqrt{[T^2]-\frac{1}{n-4}[T]^2}
    \quad\textrm{ and }\quad
    \frac{1}{n-4}[T]
\end{equation}
where it might have been necessary to extend the field of numbers in
order for the square root to exist.  It remains to find the
multiplicities of these and to identify the cases when two or all
three of them coincides.

Let the algebraic multiplicities of the eigenvalues be $a$, $b$ and
$c$.  Then we find that
\begin{equation}
    [T^2] =
    \frac{n}{4}\Bigl([T^2]-\frac{1}{n-4}[T]^2\Bigr)
    -\frac{c}{4}\Bigl([T^2]-\frac{n}{(n-4)^2}[T]^2\Bigr).
\end{equation}
Thus $c=n-4$ is always a solution to this equation.  It is the only solution
if the last parenthesis is non-zero.  Assume this is the case,
i.e., $[T^2]\not= n [T]^2/(n-4)^2$.  Note that this happens precisely
when the last eigenvalue is distinct from the others.  Calculating the
trace of $T$ one finds
\begin{equation}
    [T] = [T] + \frac{a-b}{2}\sqrt{[T^2]-\frac{1}{n-4}[T]^2}.
\end{equation}
If the square root is non-zero then $a=b$, and $c=n-4$ implies that
$a=b=2$.  In this case all eigenvalues are distinct.  If the
square root is zero, i.e., $[T^2]=[T]^2/(n-4)$, then
equation~\eref{eq:rank4} can be written as
\begin{equation}
    T^2 \Bigl(T-\frac{1}{n-4}[T] \Bigr) = 0.
\end{equation}
This gives that 0 is an eigenvalue which admits a Jordan-string of
length at most two.  The fact that $c=n-4$ gives that the algebraic
multiplicity of this eigenvalue must be four.

Consider now the case $[T^2] = n [T]^2/(n-4)^2$.  Then
equation~\eref{eq:rank4} can be written as
\begin{equation}
    \Bigl(T+\frac{1}{n-4}[T]\Bigr)
    \Bigl(T-\frac{1}{n-4}[T]\Bigr)^2
    = 0.
\end{equation}
If $[T]$ is non-zero then there are two distinct eigenvalues,
$\pm[T]/(n-4)$, where the one with $+$ admits non-trivial
Jordan-strings of length at most two.  Assume that their algebraic
multiplicities are $n-a$ and $a$.  Then
\begin{equation}
    [T] = \frac{n-2a}{n-4}[T]
\end{equation}
which implies that $a=2$.

If $[T]=0$ then equation~\eref{eq:rank4} becomes
\begin{equation}
    T^3 = 0
\end{equation}
which gives that all eigenvalues are zero and that there might be
Jordan-strings of length at most three.

\end{document}